\documentstyle[aps,twocolumn,epsf]{revtex}
\begin{document}%************************************************************
 
\title{Topological spectral correlations in 2-d disordered  
systems} 
\author{Vladimir E. Kravtsov$^1$ and Vladimir I. Yudson$^2$}
\address{ $^1$International Centre for Theoretical Physics, P.O. Box
586, 
34100 Trieste, 
Italy \\and Landau Institute for Theoretical Physics, Russian
Academy of Sciences,
2 Kosygina str., 117940, Moscow r-n, Russia\\
$^2$Institute of Spectroscopy, Russian Academy of Sciences, 142092,
Troitsk, Moscow r-n, Russia.}
\date{\today}
\maketitle 
\begin{abstract} 
It is shown that the tail in the two-level spectral correlation function
$R(s)$ for 2D disordered systems depends on the system geometry and the
boundary conditions. In particular, for
closed surfaces (with no boundary) $R(s)=-\chi/(6\pi^2\beta s^2)$, 
where $\beta=1,2$ or $4$ for the orthogonal, unitary and symplectic
ensembles, and $\chi=2(1-p)$ is the Euler characteristics of the
surface with $p$ ``handles" (holes). 
The result is valid for
$g\ll s\ll g^2$ for $\beta=1,4$ and for $g\ll s\ll g^3$ for $\beta=2$,
where $g\gg 1$ is the dimensionless conductance.
\end{abstract}
\draft

Spectral correlations in complex and disordered
systems are closely related with the basic symmetries of
a system, the nature of its dynamics and the structure of the
corresponding eigenstates. It is remarkable that there are
few universal spectral statistics which all the generic
complex and disordered systems obey in the thermodynamic (TD) limit.
According to the well known Bohigas-Giannoni-Schmidt conjecture 
\cite{BGS}, a generic chaotic system is described by the Wigner-Dyson (WD)
spectral statistics which follows from one of the three universal
Gaussian ensembles of random matrices \cite{M}. In contrast, all the
generic integrable systems obey the Poisson  spectral statistics.
Depending on the strength of disorder, spectral statistics of a
$d$-dimensional disordered system of
non-interacting particles also flow 
to one of the above universal statistics as the system
size $L$ tends to infinity.
For strong disorder and for low-dimensional systems $(d=1,2)$ where all
states are
localized, the spectral statistics in the TD limit are
Poissonian. For weakly
disordered 3D systems where all states are extended, the spectral
statistics are identical to the WD \cite{Ef} in the TD
limit. 

However, the disordered systems with $d>2$ are known to undergo the
Anderson
metal-insulator transition at a certain critical disorder. At this point
the localization (correlation) length $\xi$ diverges and the
spectral statistics should be independent of the system  size $L$,
provided
that the energy difference $\omega= s \Delta$ is measured
in units of the
mean level spacing $\Delta\propto L^{-d}$. Thus at the Anderson transition
there exists
a special fixed point, the critical spectral statistics (CSS)
\cite{Shk,KLAA}. 

An amazing property of some CSS found recently
\cite{Mon}
is the sensitivity to boundary conditions and a shape of
a sample. It will follow from the
analysis below that the basic properties of the CSS take their
``canonical" form for the {\it periodic boundary conditions} (PBC).
Other boundary conditions induce some characteristic features in the
CSS which depend on the properties of the boundary.

In order to study the qualitative role of the system geometry
for the CSS we consider the system which shows all the typical features of
the critical system and  allows for a rigorous analysis. This is
a 2D weakly disordered electron  system. 
The small parameter which allows for a
rigorous treatment
is the inverse dimensionless conductance $g^{-1}=\Delta/E_{c}\ll 1$, where
$\hbar /E_{c}=t_{D}=L^{2}/D$ is the diffusion time and $D$ is the
diffusion coefficient. For not too large
system size $L\ll \xi$, where $\xi=l \exp[g]$ or $\xi=l \exp[g^2]$ is the
localization radius in the orthogonal or unitary ensemble, respectively, 
one can neglect the
localization effects and consider the dimensionless conductance $g$ as
independent of the size $L$. At such conditions the behavior of the 2D 
system
is similar to the one in the true critical point. 

The two-level
correlation function $R(s)$ in such 2D systems  has been considered
in
Ref.\cite{KL} for the case of the periodic boundary conditions {\it only}.
It has been shown that 
the two-level correlation function R(s) is {\it exponentially small} at
$s\gg g$  if one neglects the weak-localization effects. This means
that the so called Altshuler-Shklovskii  \cite{AS} tail $R(s)= C_{d}
s^{-2+d/2}$ is {\it absent} in 2D systems. This statement is true
\cite{KLAA} for a generic critical system.

In the present Letter we consider $R(s)$ at $s\gg g$
for an arbitrary 2D surface $S$ of the  area  $A_{S}$. 
We will show that in the case where the surface has a non-transparent
boundary, the two-level correlation function has a power-law tail  
$R(s)\sim 
g^{-1/2} s^{-3/2}$. 
Moreover, we will show that even in the absence of  boundaries (e.g. for
the sphere)
the correlation function $R(s)$ still has a 
power-law tail
$R(s)=C
s^{-2}$. This tail at $s\gg g$ is of the same form as the universal WD
tail for
$1\ll s\ll g$. However
the coefficient $C$  depends on  {\it topology} of the surface
 and is
equal to zero only for the torus topology which is equivalent to the PBC. 

The two-level correlation function is defined as:
\begin{equation}
\label{TLCF}
R(\omega)=\frac{1}{\rho^2}\left[\left\langle \rho\left(
E+\frac{\omega}{2}\right)\;\rho\left( 
E-\frac{\omega}{2}\right)\right\rangle -1\right],
\end{equation}
where $\langle ...\rangle$ denote averaging over all realizations,
$\rho=\langle\rho(E) \rangle$ is the average one-electron density of
states (DOS), and
$\rho(E)$ is the DOS for a particular realization of disorder:
\begin{eqnarray}%*************************************************************%
\label{rho}%**************************************************************%
\rho(E)= S^{-1}\sum_{n}\delta (E-E_{n})
\end{eqnarray}
where $E_{n}$ are exact one-electron energy levels.
Since we are interested in the energy scale $\omega=s \Delta$ which
vanishes
in the TD limit, the average DOS is considered {\it energy-independent}.

It is remarkable that in the limit $g\gg 1$ and $s\gg 1$ the spectral
correlation function $R(s)$ of the {\it
quantum} problem can be expressed  \cite{AS} entirely in
terms of the eigenvalues $\varepsilon_{\mu}\Delta/\hbar$ of the {\it
classical} diffusion problem with the Neumann boundary conditions
which correspond to the non-transparent boundary:
\begin{equation}
\label{AS}
R(s)=\frac{1}{\pi^2 \beta}\;\Re
\sum_{\mu}\frac{1}{(\varepsilon_{\mu}-i\, s)^2},
\end{equation}
where $\beta=1,2,$ or $4$ for the orthogonal, unitary and symplectic
ensembles, respectively \cite{M}. Note that the dimensionless conductance  
$g=\varepsilon_{1}-\varepsilon_{0}$ is just the gap  between the 
lowest ($\varepsilon_{0}=0$) and the first non-zero eigenvalues.

This means that independently of the details of the electron conduction
band structure and the short-range correlated random potential,  at
$g\gg
1$ there exists a region
of
$s$ where the electron level correlations depend only on the spectrum
$\varepsilon_{\mu}$ of the
Laplace-Beltrami operator $\Delta_{{\cal G}}$ on a curved surface:
\begin{equation}
\label{LB}
\Delta_{{\cal G}}= \frac{1}{\sqrt{{\cal G}}}\,\frac{\partial}{\partial
x^{i}}
\left(\sqrt{{\cal G}}\,{\cal G}^{ij}\frac{\partial}{\partial x^{j}} 
\right),
\end{equation}
where ${\cal G}^{ij}=[\hat{{\cal G}}^{-1}]_{ij}$, ${\cal G}=\det \hat{
{\cal G}}$, and $\hat{{\cal G}}$ is a metric tensor on the surface.

Eq.(\ref{AS}) can be rewritten in the following form \cite{AIS}:
\begin{equation}
\label{AIS}
K(t)=\int_{-\infty}^{+\infty}\frac{ds}{2\pi}\,
e^{-its}\,R(s)=\frac{1}{2\pi^2 \beta}\,|t|\,p(|t|),
\end{equation}
if one introduces the classical return probability $P(t,{\bf r})$
 to the
point ${\bf r}$:
\begin{equation}
\label{rp}
P(t,{\bf r})=\sum_{\mu} \exp[-t\varepsilon_{\mu}] \;[\Phi_{\mu}({\bf
r})]^2,
\end{equation} 
and the averaged return probability:
\begin{equation}
\label{avrp}
p(t)= \int_{S}d{\bf r} P(t,{\bf
r})= \sum_{\mu} \exp[-t\varepsilon_{\mu}],
\end{equation}
where $\Phi_{\mu}({\bf
r})$ is an eigenfunction of the diffusion (Laplace-Beltrami) operator
and $t=T/t_{H}=T\Delta/\hbar$ is time in units of the Heisenberg time
$t_{H}$.

Eq.(\ref{AIS}) sets the relationship between the tail in $R(s)$ at $s\gg
1$ and the small-time behavior of the return probability
$p(t)$ at $t\ll 1$. There are two  regions of $t$ with different 
behavior of $p(t)$: the diffusion
region $t\ll 1/g$ and the ergodic region $1/g\ll t\ll 1$. 
In the ergodic region
$t\varepsilon_{\mu}\gg 1$ for all $\mu\neq 0$, so that only the zero mode
$\varepsilon_{0}=0$ contributes to Eqs.(\ref{rp}),(\ref{avrp}). In this
region $p(t)=1$ is independent of $t$, and one obtains the universal WD
result $R(s)= -\frac{1}{\pi^2\beta}\,s^{-2}$. 

In the diffusion region
the main contribution to the return probability is given by the
continuous approximation in Eq.(\ref{avrp}). In this approximation 
$p(t)$ is well known to be proportional
to $1/t$ in 2D systems thus giving rise to a constant
$K(t)$.
It is the property of all the critical
systems that the main contribution to $K(t)$ is independent of $t$ for
sufficiently small times \cite{AKL,CKL}. This means that the tail
in
R(s) for $s\gg g$ is determined in critical systems entirely by the
{\it corrections} to the return probability. There are two types of
such corrections. One source of corrections is the {\it weak-localization}
effects considered in Ref.\cite{KL} or the  finite-size corrections to
scaling considered in Ref.\cite{AKL}.
In 2D systems they lead to: 
\begin{equation}%***********************************************************%
\label{WLC}%*************************************************************%
R_{WL}(s)\sim
\left\{\matrix{
\mp g^{-2}s^{-1},
   &  \beta=1,4; \cr
    -g^{-3} s^{-1},  &  \beta=2 \cr} \right.
\end{equation}
Another source of corrections which has not been considered so far is 
the corrections to the continuous approximation in Eq.(\ref{avrp})
that is the difference between the sum and the corresponding integral.
This correction is sensitive to the {\it boundary conditions} and 
{\it topology} of the surface.

The sensitivity of $p(t)$ to the boundary conditions originates from the
${\bf
r}$-dependence of the return probability $P(t,{\bf r})$. 
If the point ${\bf r}$ is at a sufficiently small distance $x\ll L$
from a smooth boundary, the ${\bf r}$-dependent correction $\delta
G(t,{\bf r},{\bf r})=\delta P_{b}(t,{\bf r})$ to the Green's
function of the diffusion equation
with the Neumann boundary
conditions, is given by the ``image source":  
$\delta P_{b}(t,{\bf r})=(4\pi DT)^{-1}\exp[-x^2/DT]$.  Then
assuming the smooth boundary of the length ${\cal L}$ one obtains
\cite{BH} for the
corresponding correction to the {\it averaged} return probability $p(t)$:
\begin{equation}
\label{bef}
\delta p_{b}(t)=\int_{0}^{\infty}\delta P_{b}(t,{\bf r})\,{\cal L}dx= 
\frac{{\cal L}}{8\sqrt{\pi DT}}.
\end{equation}
The boundary-induced tail in $R(s)$ is found immediately from 
Eqs.(\ref{AIS}),(\ref{bef}):
\begin{equation}
\label{bound}
R_{b}(s)= -
\frac{1}{8\beta\,(\pi g_{b})^{1/2}\,(2\pi s)^{3/2}}, 
\end{equation}
where $g_{b}=D/{\cal L}^2 \Delta$. Note that for ${\cal L}\sim L$ we have
$g_{b}\sim g$.

Comparing Eqs.(\ref{WLC}),(\ref{bound}) one can see that the
boundary-induced power-law tail
in $R(s)$ is larger than the 
localization-induced tail  if  $g\ll s\ll g^3$ for $\beta=1,4$
and $g\ll s\ll g^5$ for $\beta=2$. 

If the boundary is absent at all ($R_{b}(s)\equiv 0$), still there is a
correction to the continuous approximation in Eq.(\ref{avrp}).
This correction is {\it universal} and depends only on the {\it topology}
of the surface. In order to understand how such a topological correction
may arise at times $T$ which are small compared with the diffusion time,
let us consider
the return probability $P(t,{\bf r})$. For times $T\ll t_{D}$ ($t\ll 1/g$)
the diffusing electron
probes only the vicinity of the starting point ${\bf r}$. Therefore, the
correction to the return probability $P(t,{\bf r})$ may depend only on the
{\it local} curvature. We will show below (see Eq.(\ref{res})) that it is
proportional to
the Gauss curvature 
$K_{\bf r}=k^{(1)}_{{\bf r}}\,k^{(2)}_{{\bf r}}$ in a
point
${\bf r}$, where $k^{(1)}$ and $k^{(2)}$ are the principal curvatures. 
An information about topology of a surface arises
because
of
the integration over all positions of the starting point ${\bf r}$ in
Eq.(\ref{avrp}). Indeed, according to the Gauss-Bonnet theorem \cite{DNF}
the integral of $K_{{\bf r}}$ over a surface $S$ is related with the Euler
characteristic $\chi(S)$ of the surface:
\begin{equation}
\label{GB}
\int_{S} K_{{\bf r}} \,d\sigma = 2\pi \chi(S).
\end{equation}
The quantity $\chi(S)=V+F-E$ is related with the number of vertices $V$,
edges
$E$ and
faces $F$ of the surface triangulation and depends on the
{\it
connectivity} of the surface, i.e. on the number $p$ of ``handles"
(holes):
\begin{equation}
\label{con}
\chi(S)=2(1-p),
\end{equation}
where $p=0$
for a sphere and $p=1$ for a torus.

In the simplest case of a sphere where each
eigenvalue $\varepsilon_{\mu}=l(l+1)$ of the Laplace-Beltrami operator
Eq.(\ref{LB}) is $(2l+1)$ times degenerate, the topological correction to
$p(t)$ at $tg\ll 1$ can be obtained by a straightforward summation in
Eq.(\ref{avrp}) using the well known \cite{KK} formula
$\sum_{0}^{\infty}f(l)\approx\int_{-1/2}^{\infty}f(l) dl +
\frac{1}{24}f^{'}(-1/2)$: 
\begin{equation}
\label{ss}
p(t)=\sum_{l=0}^{\infty} (2l+1)\, \exp[-tg l(l+1)]\approx
\frac{1}{tg}+\frac{1}{3}.
\end{equation}

One can see that for a sphere ($p=0$) the topological correction to $p(t)$
at $tg\ll 1$ is
a {\it universal constant} $\frac{1}{3}$. Then using Eq.(\ref{con})
and $\delta p(t)\propto \chi(S)$ we conclude that for a generic surface
the topological correction reads:
\begin{equation}
\label{topcor}
\delta p_{top}(t)=\frac{1-p}{3}.
\end{equation} 
This correction is time-independent and one immediately obtains from 
Eq.(\ref{AIS}) a topological tail in $R(s)$: 
\begin{equation}
\label{tt}
R_{top}(s)= -\frac{1-p}{3\pi^2\beta s^2},\;\;\;\;\;\;(s\gg g).
\end{equation}
Comparing this correction with the weak-localization correction
Eq.(\ref{WLC}) we see that the latter is smaller at $g\ll s \ll g^2$
or $g\ll s \ll g^3$ for $\beta=1,4$ or $\beta=2$, respectively.
Note that the topological tail has the same form as the universal WD tail.
However, the former is valid in the {\it diffusion} region $s\gg g$ while
the latter is valid in the {\it ergodic} region $s\ll g$.
 
The result Eq.(\ref{topcor}) follows from the theory of the Laplacian on
Riemannian manifolds (TLRM)
\cite{HKT} and has been used \cite{BH} to find corrections to the
semiclassical density of states in quantum billiards and resonators
\cite{corr}.
Here we present an elementary derivation of Eq.(\ref{topcor})
which does not require the full power of the TLRM. 

For an arbitrary non-singular point ${\bf r}$ on a surface, it is always
possible to choose the local system of co-ordinates
$(\xi_{1},\xi_{2})$  such that   the inverse metric tensor ${\cal
G}^{ij}(\xi_{1},\xi_{2})$ in the vicinity of the origin (point ${\bf r}$)
takes the form:
\begin{equation}
\label{MT}
{\cal G}^{11}={\cal G}^{22}=1,\;\;\; {\cal G}^{12}={\cal
G}^{21}=-K_{{\bf r}}\xi_{1}\xi_{2},
\end{equation}
where $K_{{\bf r}}$ is the  Gaussian curvature at a point ${\bf r}$, and
higher order terms in $\xi^{1,2}$ are omitted. Then
we
obtain for the Laplace-Beltrami
operator Eq.(\ref{LB}):
\begin{eqnarray}
\label{LBp}
&&\frac{\partial^2}{\partial \xi_{1}^2}+\frac{\partial^2}
{\partial \xi_{2}^2}-K_{{\bf r}}\left(\frac{\partial}{\partial
\xi_{1}}\xi_{1}\xi_{2}
\frac{\partial}{\partial \xi_{2}}+\frac{\partial}{\partial \xi_{2}}
\xi_{1}\xi_{2}\frac{\partial}{\partial \xi_{1}} \right) \nonumber
\\ &\equiv&
\hat{\Delta} +\hat{V}.
\end{eqnarray}
As has been mentioned above, for small times $T\ll t_{D}$ the term
proportional to the
Gaussian curvature $K_{\bf r}$ can be treated as perturbation. The return
probability $P(t,{\bf r})$ is given by the Green's function $G(t,{\bf
r},{\bf 
r})$
of the diffusion operator $\partial/\partial \tilde{t}-\Delta_{{\cal
G}}$, where $\tilde{t}=DT$. Then all we should do in order to find the
correction to the local return probability is to compute the correction to
the
Green's function $G^{(1)}=G^{(0)}\hat{V}G^{(0)}$, where $G^{(0)}=(4\pi
\tilde{t})^{-1}\,\exp[-(\xi-\xi')^{2}/4\tilde{t}]$.
The result is:
\begin{equation}
\label{res}
P(t,{\bf r})=\frac{A_{S}}{4\pi \tilde{t}}+\frac{K_{{\bf r}}}{12\pi},
\end{equation}
where $A_{S}$ is the surface area. The first term in this
expression corresponds to the continuous approximation.
The second one is the correction due to the local curvature $K_{{\bf r}}$.
This very term has been used in deriving Eq.(\ref{topcor}).

In conclusion, we have considered the power-law tail in the
two-level correlation function R(s) for 2D disordered systems with the
diffusion motion of electrons. In the limit of large dimensionless
conductance $g\gg 1$, there is an interval of $s\gg g$ where the
tail is entirely determined by the geometry of the sample.
It consists of the two contributions: a boundary contribution given by
Eq.(\ref{bound}), and a topological contribution given by Eq.(\ref{tt})
which are the main results of the Letter. For closed surfaces without a
boundary, the tail in $R(s)$ for $s\gg g$ is determined only by topology
and is of the
same form as the universal
Wigner-Dyson tail for $1\ll s\ll g$. However, the numerical topological
pre-factor depends
on the connectivity of the surface and is negative for  single-connected
surfaces (sphere), zero for  surfaces of the torus topology, and {\it
positive} for surfaces of higher genus.
\acknowledgments
We are grateful to B.L.Altshuler, E.Bogomolny, I.V.Lerner, A.D.Mirlin,
G.Montambaux,
and
I.V.Yur-\\kevich for stimulating discussions. The work was supported by
INTAS-RFBR grant No.95-0675, RFBR grant No.96-02-17133, and the US CRDF
Award No.RP1-209 (V.E.K.) and RFBR grant No.98-02-16062 (V.I.Y.).
V.I.Yudson acknowledges the hospitality of the ICTP in Trieste, where
the work has been completed.


\begin{references}
\bibitem{corr}
Note that in Ref.\cite{BH} the expression for $\delta p_{top}(t)$ similar
to Eq.(\ref{topcor}) has an incorrect extra factor of 1/2 that has been
rewritten in a number of further applications to quantum billiards.
\end{references}
\end{document}